\documentclass[floatfix,reprint,superscriptaddress,footinbib,aip,cha]{revtex4-1}
\pdfoutput=1 

\usepackage[utf8]{inputenc}
\usepackage[english]{babel}

\usepackage{amsmath}
\usepackage{amsfonts}
\usepackage{amssymb}
\usepackage[normalem]{ulem}
\usepackage[hyperindex,colorlinks,hyperfootnotes = false,citecolor = blue,]{hyperref}
\usepackage[usenames]{color}

\usepackage{graphicx}

\newcommand{\LW}[2]{\mathcal{L}_{#1 (B^T #2)}} 
\newcommand{\LpiS}{L^{\pi_\sigma} \,}
\newcommand*{\defeq}{\mathrel{\vcenter{\baselineskip0.5ex \lineskiplimit0pt
                     \hbox{\scriptsize.}\hbox{\scriptsize.}}}=}
\newcommand*{\defeqr}{=\mathrel{\vcenter{\baselineskip0.5ex \lineskiplimit0pt
                     \hbox{\scriptsize.}\hbox{\scriptsize.}}}}

\newcounter{remarkcounter}

\begin{document}
\title{Graph partitions and cluster synchronization in networks of oscillators}
\author{Michael T. Schaub}
\email{michael.schaub@uclouvain.be}
\affiliation{ICTEAM, Universit\'e catholique de Louvain,  B-1348 Louvain-la-Neuve, Belgium} 
\affiliation{naXys and Department of Mathematics, University of Namur,  B-5000 Namur, Belgium}
\author{Neave O'Clery}
\affiliation{Center for International Development, Harvard University, Cambridge, MA 02138, United States of America}
\author{Yazan N. Billeh}
\affiliation{Computation and Neural Systems Program, California Institute of Technology, Pasadena, CA 91115, United States of America}
\author{Jean-Charles Delvenne}
\affiliation{ICTEAM, Universit\'e catholique de Louvain,  B-1348 Louvain-la-Neuve, Belgium} 
\affiliation{CORE, Universit\'e catholique de Louvain,  B-1348 Louvain-la-Neuve, Belgium} 
\author{Renaud Lambiotte}
\affiliation{naXys and Department of Mathematics, University of Namur,  B-5000 Namur, Belgium}
\author{Mauricio Barahona}
\email{m.barahona@imperial.ac.uk}
\affiliation{Department of Mathematics, Imperial College London, London SW7 2AZ, United Kingdom}

\begin{abstract}
    Synchronization over networks depends strongly on the structure of the coupling between the oscillators. 
    When the coupling presents certain regularities, the dynamics can be coarse-grained into clusters by means of External Equitable Partitions of the network graph and their associated quotient graphs. 
 We exploit this graph-theoretical concept to study the phenomenon of cluster synchronization, in which different groups of nodes converge to distinct behaviors.  
    We derive conditions and properties of networks in which such clustered behavior emerges, and show that the ensuing dynamics is the result of the localization of the eigenvectors of the associated graph Laplacians linked to the existence of invariant subspaces. 
    The framework is applied to both linear and non-linear models, first for the standard case of networks with positive edges, before being generalized to the case of signed networks with both positive and negative interactions. We illustrate our results with examples of both signed and unsigned graphs for consensus dynamics and for partial synchronization of oscillator networks under the master stability function as well as Kuramoto oscillators. 
\end{abstract}
\maketitle

\begin{quotation}
Synchronization of coupled oscillators is ubiquitous in nature: from the rhythmic flashing of fireflies or the orchestrated chirping of crickets to the entrainment of circadian rhythms or the coherent firing of neurons in epilepsy to the dynamics of man-made networks, such as power grids and computer networks. 
Synchronization is also related to consensus processes, such as the flocking of birds or shoaling of fish, or opinion formation in social networks.

Previous studies have typically focused on complete synchronization, where all agents on a network converge to the same dynamics.  
However, many networks display patterns of synchronized clusters, where different groups of agents converge to distinct behaviors. 
Here we use tools from graph theory to study the phenomenon of cluster synchronization. 
We show that cluster synchronization can emerge in networks that can be partitioned into groups according to an \emph{external equitable partition} (EEP) of the graph.
Our graph-theoretical approach allows us to extend the analysis to networks with positive and negative links, which are important to describe social interactions and inhibitory-excitatory interactions in biology.     
We showcase applications to consensus dynamics and to generic synchronization of oscillators, including the classic Kuramoto model, and discuss general applications to networked systems of interacting agents. \end{quotation}

\section{Introduction}
Synchronization phenomena are prevalent in networked systems in biology, physics, chemistry, as well as in social and technological networks.
The study of these pervasive processes thus spans many disciplines leading to a rich literature on this subject~\cite{Strogatz2000,Boccaletti2002,Arenas2008,Doerfler2014,Nishikawa2015,Panaggio2015,Rodrigues2016}. 
The synchronization literature has traditionally focused on the problem of total synchronization, initially under mean field or global coupling~\cite{Kuramoto2012,Strogatz2000} and more recently studying how total synchronization relates to properties of the interaction topology and the dynamics of the individual agents~\cite{Strogatz2000,Boccaletti2002,Arenas2008,Doerfler2014,Jadbabaie2003,Heagy1994,Pecora1998,Barahona2002,pecora2005synchronization,Jadbabaie2004, Arenas2008}.  

Currently, there is a surge of interest in localized synchronization processes, where parts of the network become locally synchronized. This phenomenon may also be referred to as \emph{partial synchronization}, \emph{cluster synchronization}, or \emph{polysynchrony}~\cite{Judd2013,DHuys2008,Lu2010,Dahms2012,Do2012,Fu2013,Rosin2013,Sorrentino2007,Williams2013,Pecora2014,Sorrentino2016}.
Recent work has shown that the predisposition of a network of coupled oscillators to exhibit cluster synchronization is intimately linked to symmetries present in the coupling~\cite{Pecora2014,Sorrentino2016}.
In particular, Pecora and collaborators showed how one can use the inherent symmetry group of the network to block-diagonalize the coupling, thereby assessing the stability of cluster synchronization under the master stability function (MSF) formalism~\cite{Heagy1994,Pecora1998}.

Here, we will also be concerned with the subject of cluster synchronization 
of oscillators in networks with general topologies.
However, instead of using a group-theoretic viewpoint, we will consider this problem from an alternative 
\emph{graph-theoretical} perspective. Specifically, we derive results for cluster synchronization in networks of oscillators using the notion of \textit{external equitable partition} (EEP), a concept that has gained prominence in systems theory to study consensus processes~\cite{Egerstedt2012,OClery2013, cardoso2007laplacian,martini2010controllability}.
The use of EEPs emphasizes the presence of an \textit{invariant subspace} in the coupling structure and
leads to a coarse-grained description of the network in terms of a quotient graph. This approach complements the group-theoretical symmetry viewpoint in~Refs.~\cite{Pecora2014,Sorrentino2016}, while
also encompassing the analysis of networks of Kuramoto oscillators~\cite{Kuramoto1975,Kuramoto2012}, a prototypical model for phase synchronization 
which does not lend itself to the MSF formalism.

In addition, we show how the EEP perspective of cluster synchronization 
can be generalized to \emph{signed networks}, 
i.e., graphs with links of positive and negative weights.
To do this, we define the notion of \emph{signed external equitable partition} 
(sEEP) and demonstrate its applicability on \emph{structurally balanced signed networks},
a classic model from the theory of social networks~\cite{Heider1946,Cartwright1956}. 
In structurally balanced signed networks, linear consensus dynamics leads to a form of `bipolar consensus'~\cite{Altafini2013,Altafini2013a}, in which nodes split into two factions, i.e., nodes inside the same faction converge to a common value while the other faction converges to the same value with opposite sign. 
In the synchronization setting, we demonstrate that the presence of sEEPs can induce a \textit{bipolar cluster synchronization}, in which each group of oscillators may be divided into two `out-of-phase' groups, with trajectories of equal magnitude but opposite sign.  
Below, we show how these results appear for signed networks under the MSF framework as well as for Kuramoto oscillators.  
\\

\paragraph*{\textbf{Notation:}}
Our notation is standard. 
The number of nodes (vertices) in the network is denoted by $N$; the number of edges (links) by $E$.
We denote the adjacency matrix of the graph by $A=A^T$, where $A_{ij}$ corresponds to the weight of the coupling between node (oscillator) $i$ and $j$.
The graph Laplacian matrix is defined as ${L=D-A}$, where $D=\text{diag}(A\mathbf{1})$ is the matrix containing the total coupling strength of each node on the diagonal, i.e., ${D_{ii} =  \sum_{j}A_{ij}}$.
From this definition, it is straightforward to see that the vector of ones $\mathbf{1}$ 
is an eigenvector of $L$ with eigenvalue $0$.
It is well known that the Laplacian may be decomposed as $L= BWB^T$, where $B$ is the node-to-edge incidence matrix and $W$ is a diagonal matrix containing the (positive) weights of the edges.
It therefore follows that the Laplacian is a positive semidefinite matrix.

To simplify notation, but without loss of generality, our exposition below is presented for unweighted graphs, i.e., $W=I$. However, all our results apply to weighted graphs by using edge weight matrices appropriately.  

\section{External equitable partitions}

External equitable partitions are of interest because the existence of an EEP in a graph has 
implications for its spectral properties and, consequently, for dynamical processes associated with 
the graph. EEPs extend the notion of equitable partition (EP).  
An EP splits the graph into non-overlapping cells $\{\mathcal C_i\}$ (groups of nodes), such that the number of connections to cell $\mathcal C_j$ from any node $v\in\mathcal C_i$ is only dependent on $i,j$.  Stated differently, the nodes inside each cell of an EP have the same out-degree pattern with respect to every cell. 
For EEPs, this requirement is relaxed so that it needs to hold only for the number of connections between \textit{different} cells $\mathcal C_i, \mathcal C_j$ ($i\neq j$).

Algebraically, these definitions can be represented as follows~\cite{cardoso2007laplacian,Egerstedt2012,Chan1997}.
A partition of a graph with $N$ nodes into $c$ cells is encoded by the $N \times c$ indicator matrix $H$: 
$H_{ij} =1$ if node $i$ is part of cell $\mathcal C_j$ and $H_{ij}=0$ otherwise.
Hence the columns of $H$ are indicator vectors $\mathbf{h}_i$ of the cells: 
\begin{align}
\label{eq:H_cols}
H & \defeq [\mathbf{h}_1,\ldots,\mathbf{h}_c].
\end{align}

Given the Laplacian matrix $L$ of a graph, we can write the definition of an EEP as follows:
\begin{equation}
\label{eq:EEP_def}
    LH = HL^\pi.
\end{equation}
Here $L^\pi$ is the $c \times c$ Laplacian of the quotient graph induced by $H$: 
\begin{equation}
\label{eq:Lpi}
    L^\pi =  (H^TH)^{-1}H^T  LH = H^+LH,
\end{equation}
where the $c \times N$ matrix $H^+$ is the (left) Moore-Penrose pseudoinverse of $H$. 
Observe that multiplying a vector $\mathbf{x} \in \mathbb R^N$ by $H^T$ from the left sums up the components within each cell, and that $H^TH$ is a diagonal matrix with the number of nodes per cell on the diagonal. Hence $H^+$ may be interpreted as a \emph{cell averaging operator}~\cite{OClery2013}. 

\begin{figure}
\includegraphics[width=.45 \textwidth]{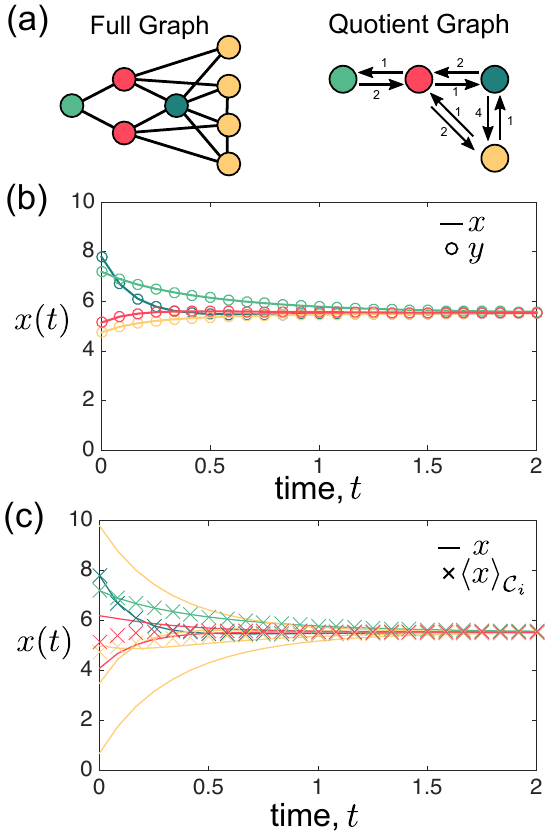}
\caption{\textbf{External equitable partitions and invariant consensus dynamics.} 
    \textbf{(a)} A graph with $N=8$ nodes with an external equitable partition into four cells (indicated with colors) and its associated quotient graph.  
    \textbf{(b)} The evolution of the consensus dynamics on the full graph~\eqref{eq:consensus} 
    from an initial condition $\mathbf{x}=H\mathbf{y}$ is shown with solid lines.  The associated quotient 
    dynamics~\eqref{eq:invariant_cells} governing $\mathbf{y}$ is shown with circles. Once all states within each cell are equal (i.e., they are cluster-synchronized), the dynamics will remain cluster-synchronized and its dynamics will be described by the quotient dynamics for all times.
    \textbf{(c)} For consensus dynamics, the quotient graph dynamics also describes the cell-averaged dynamics (crosses) of the unsynchronized full graph dynamics (solid lines), as given by~\eqref{eq:cell_averages}.}
\label{fig:schematic_EEP}
\end{figure}

The quotient graph associated with an EEP is a coarse-grained version of the original graph, such
that each cell of the partition becomes a new node and the weights between 
these new nodes are the out-degrees between the cells in the original graph 
(see Fig.~\ref{fig:schematic_EEP}a).
Although the Laplacian of the original graph is symmetric, 
the quotient Laplacian will be asymmetric in general.
Note that, from the definition of the Laplacian, there is always a trivial EEP 
in which the whole graph is grouped into one cell, i.e.,  $H = \mathbf{1}$ and $L^\pi = 0$.

From~\eqref{eq:EEP_def}--\eqref{eq:Lpi}, 
the definition of the EEP can be rewritten solely in terms of $L$:
\begin{equation}\label{eq:subspace_id}
    LH = HH^+LH = P_H LH,
\end{equation}
where $P_H \defeq HH^+$ is the projection operator onto the cell subspace,
i.e., it defines an orthogonal projection onto the range of $H$.

The operator $P_H$ commutes with~$L$:
\begin{equation}
\label{eq:PH_commutes}
    LP_H = H L^\pi H^+ = P_HLP_H =  P_HL,
\end{equation}
which follows from~\eqref{eq:EEP_def}, \eqref{eq:subspace_id} and the symmetry of $L$ and $P_H$.  
Using the commutation~\eqref{eq:PH_commutes}, it is easy to show that:
\begin{equation}
\label{eq:H+_commutator}
  H^+L =  L^\pi H^+,
\end{equation}
which summarises the relationship between the cell averaging operator $H^+$ and the Laplacians of
the original and quotient graphs.\\

\refstepcounter{remarkcounter}
\noindent \textbf{\emph{Remark \theremarkcounter}} \textbf{\emph{[Equitable partitions and coupling via
adjacency matrices]: }}
\label{rem_equitable} 
\textit{
It is instructive to consider EEPs with respect to the stricter requirement of equitable partitions (EPs).
Given the adjacency matrix $A$ of a graph, an equitable partition encoded by the indicator matrix $H_\mathrm{EP}$ must fulfil:
\begin{equation}
    \label{eq:EP_def}
    AH_\mathrm{EP} = H_\mathrm{EP} A^{\pi_{*}}.
\end{equation}
Hence we can define the adjacency matrix of the EP quotient graph $A^{\pi_{*}}$ 
induced by $H_\mathrm{EP}$ as:
\begin{equation}
    A^{\pi_{*}} = H^+_\mathrm{EP}\, A \,H_\mathrm{EP}.
\end{equation}
The adjacency matrix $A^{\pi_*}$ has diagonal entries corresponding to self-loops in the quotient graph of the EP, reflecting the number of edges between any two nodes inside each cell.
In contrast, the adjacency matrix of an EEP cannot be uniquely defined; thus, such in-cell information is not consistently specified.
On the other hand, the quotient graphs of both EPs and EEPs have consistently defined Laplacian matrices, due to the well-known invariance of the Laplacian to the addition of self-loops in a graph,
so that the quotient Laplacian is unaffected by the internal connectivity inside each cell. 
This algebraic argument clarifies why EEPs are defined in terms of the Laplacian~\eqref{eq:EEP_def}. 
It also follows directly that every EP is necessarily an EEP, while the converse is not true.
}
\\

\refstepcounter{remarkcounter}
\noindent \textbf{\emph{Remark \theremarkcounter}} \textbf{\emph{[Network symmetries and (external) equitable partitions]: }}
\label{rem_network_symmetries} 
\textit{
Recently, Pecora \textit{et al.}~\cite{Pecora2014,Sorrentino2016} used the symmetry groups of a graph and their associated irreducible representations to identify possible synchronization clusters in networks of oscillators and to assess their stability.
Their group-theoretical analysis is intimately related to the graph-theoretical perspective presented here.
Indeed, the symmetry groups of the graph induce orbit partitions.
Every orbit partition is an equitable partition, yet the converse is not true: 
there exist EPs not induced by any symmetry group~\cite{Kudose2009,Chan1997}.
Recall that EEPs are a relaxation of EPs in the sense that EEPs disregard the connections inside each cell. Consequently, EEPs are defined in terms of the Laplacian matrix~\eqref{eq:EEP_def}, in contrast to EPs being defined in terms of the adjacency matrix~\eqref{eq:EP_def}. 
Interestingly, recent work of~\citet{Sorrentino2016} introduced `adjusted orbit partitions' induced by symmetry groups of a ``dynamically equivalent coupling matrix'' in which internal connections inside each cluster are ignored. 
Such `adjusted orbit partitions' are in fact EEPs but, as for EPs, there exist EEPs that cannot be generated by the symmetry groups of such dynamically equivalent coupling matrices.
In this sense, EEPs provide a generalised setting that includes the group-theoretical orbit
partitions as a particular case.  }

As EEPs are a larger class of partitions than EPs and Laplacians are of wide interest in applications, we concentrate here on networks with Laplacian coupling.
All our results can be applied straightforwardly to systems in which the coupling is described by the adjacency matrix, by considering EPs rather than EEPs.

\section{Cluster synchronization under the external equitable partition}  

We first use EEPs to study cluster synchronization on standard networks, i.e., defined by connected undirected graphs with positive weights. 
We start by considering results for linear consensus, and then apply the framework to  nonlinear cluster synchronization both under the MSF formalism as well as Kuramoto networks.

\subsection{Dynamical implications of EEPs: the linear case}
The definition of the EEP~\eqref{eq:EEP_def} can be understood as a `quasi-commutation' relation, which signals a certain invariance of the partition encoded by $H$ with respect to the Laplacian $L$. Similarly, Eq.~\eqref{eq:H+_commutator} shows that the cell averaging operator $H^+$ exhibits a (distinct) invariance with respect to $L$. 
In particular, Eq.~\eqref{eq:EEP_def} implies that the associated cell indicator matrix 
$H$ spans an invariant subspace of $L$, whence it follows that there exists a set of eigenvectors which is localised on the cells of the partition.
Furthermore, the eigenvalues associated with the eigenvectors spanning the invariant subspace are shared with 
$L^\pi$, the Laplacian of the quotient graph~\cite{OClery2013}.
If $L$ has degenerate eigenvalues, an eigenbasis can still be chosen so that it is localised on the cells
of the partition~\cite{OClery2013}.

The properties of the EEP~\eqref{eq:EEP_def}--\eqref{eq:H+_commutator} have noteworthy consequences for linear dynamics dictated by $L$, as illustrated by the case of linear consensus dynamics~\cite{OClery2013}:
\begin{equation}\label{eq:consensus}
    \dot{\mathbf{x}} = -L\mathbf{x},
\end{equation}
where the $N \times 1$ vector $\mathbf{x}$ describes the state of the system.

First, as shown in Fig.~\ref{fig:schematic_EEP}b, the EEP is consistent with a form of invariance akin to `cluster consensus'. 
In particular, if the initial state vector is given by $\mathbf{x} = H \mathbf{y}$ for some arbitrary $\mathbf{y}$ (i.e., all the nodes within cell $\mathcal{C}_i$ have the same value $y_i$), the nodes inside the cells remain identical for all times and their dynamics is governed by the quotient graph:
\begin{align}
\label{eq:invariant_cells}
    \dot{\mathbf{x}} & = H\dot{\mathbf{y}} \quad \text{where} \quad     
\dot{\mathbf{y}}  = - L^\pi \mathbf{y}.
\end{align}
This follows directly from $LH\mathbf{y} = HL^\pi \mathbf{y}$. 

Second, the dynamics of the cell-averaged states 
$\langle  \mathbf{x} \rangle_{\mathcal{C}_i} $
is governed by the quotient graph:
\begin{equation}
\label{eq:cell_averages}
  \frac{d  \langle \mathbf{x} \rangle_{\mathcal{C}_i}}{dt} = 
    -L^\pi \langle \mathbf{x} \rangle_{\mathcal{C}_i} \quad \text{where} \quad {\langle  \mathbf{x} \rangle_{\mathcal{C}_i} \defeq  H^+  \mathbf{x}},
\end{equation}
which follows from $H^+L\mathbf{x} = L^\pi H^+ \mathbf{x}$.
Thus, the cell-averaged dynamics is governed by a lower dimensional linear model, with
dimensionality equal to the number of cells in the EEP (see Fig.~\ref{fig:schematic_EEP}c).

Third, the results obtained for the autonomous dynamics with no inputs~\eqref{eq:consensus}  
can be equivalently rephrased for the system with a bounded input $\mathbf{u}(t)$: 
\begin{equation}\label{eq:consensus_input}
    \dot{\mathbf{x}} =-L\mathbf{x} + \mathbf{u}(t).
\end{equation}
In particular, similarly to~\eqref{eq:invariant_cells}, we also have
cell invariance under inputs:
if we apply an input consistent with the cells of an EEP 
(i.e., ${\mathbf{u}(t) = H\mathbf{v}(t)}, {\mathbf{v}(t)\in \mathbb R^c}$), 
the nodes inside each cell remain identical for all times~\cite{OClery2013}. 
This simple insight, which follows from the impulse-response of the 
linear system~\eqref{eq:consensus_input}, will be useful 
when analysing nonlinear synchronization protocols. 

Finally, it is important to remark that since there is always a trivial EEP spanning the complete graph 
(with $H = \mathbf{1}$), all the results above and henceforth can be trivially applied to the case of global consensus (global synchronization) as a particular case.

\subsection{EEPs and nonlinear cluster synchronization within the MSF framework}
\label{sec:MSFetc}

We now extend the notions introduced above to a more general setting describing the dynamics of interconnected nonlinear systems.
This framework is known as the Master Stability Function (MSF) and has been pioneered by Pecora and co-workers~\cite{Heagy1994,Pecora1998,Barahona2002}.

We consider networks of identical coupled oscillatory nonlinear systems in which the dynamics of each node $i$ is described by:
\begin{align}\label{eq:coupled_oscillators_index}
    \dot{\mathbf{x}}_i &=\mathbf{F}(\mathbf{x}_i) - \gamma \sum_j L_{ij} \mathbf{G}(\mathbf{x}_j),
\end{align}
where $\gamma$ is a parameter that regulates the coupling strength; ${\bf x}_i \in \mathbb{R}^d$ is the state vector of node $i$; ${\mathbf{F}:\mathbb{R}^d \rightarrow \mathbb{R}^d}$ is the intrinsic dynamics of each node; and the coupling function $\mathbf{G}:\mathbb{R}^{d} \to \mathbb{R}^d$ specifies how the nodes in the network interact according to the interconnection topology described by the graph Laplacian $L$. 

Although, as discussed above, we could consider a coupling mediated by the adjacency matrix (and associated EPs), we concentrate here on the case of Laplacian coupling (and associated EEPs) as the more generic case of interest in the literature . 

To facilitate the subsequent discussion, we define 
${\mathbf{x} \defeq [{\bf x}_1^T,...,{\bf x}_N^T]^T \in \mathbb{R}^{N d}}$ and 
use the Kronecker product to rewrite~\eqref{eq:coupled_oscillators_index} compactly 
as a $Nd$-dimensional system of ODEs:
\begin{flalign}
\label{sync}
\dot{\bf x}&=\mathbf{F}_N({\bf x}) - \gamma (L \otimes I_d) {\bf G}_N({\bf x})
\end{flalign}
where $\mathbf{F}_N(\mathbf{x}) \defeq [{\bf F( x}_1)^T,\ldots,{\bf F(x}_N)^T]^T \in \mathbb{R}^{N d}$, 
${\bf G}_N({\bf x}) \defeq [{\bf G( x}_1)^T,\ldots,{\bf G(x}_N)^T]^T\in \mathbb{R}^{Nd}$ and $I_d$ is the
$d$-dimensional identity matrix.

A cluster-synchronized state consistent with an EEP with indicator matrix $H$ is then given by:
\begin{align}
\label{eq:cluster_state}
\mathbf{x}_s (t) &= (H\otimes I_d) \, \mathbf{y}_s(t) \quad \text{where} \\
\mathbf{y}_s(t) & = [\mathbf{s}_1(t)^T,\ldots, \mathbf{s}_c(t)^T]^T\in \mathbb R^{cd}.
\end{align}

\begin{figure}[t!]
\centering
\includegraphics[width=.45 \textwidth]{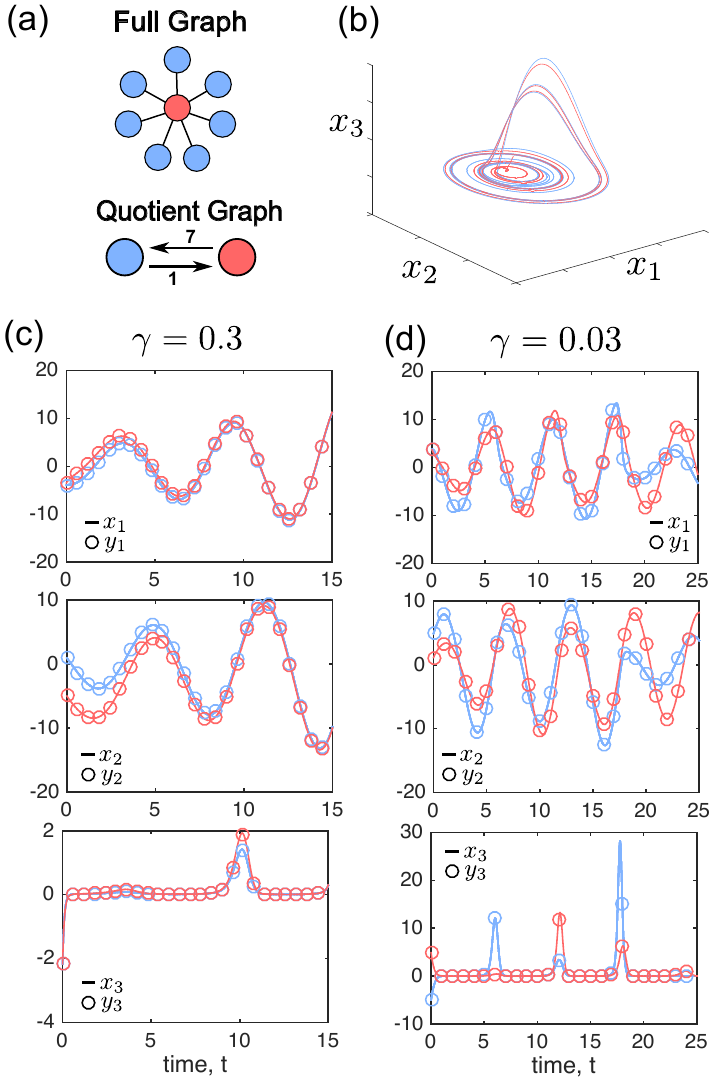}
\caption{\textbf{Synchronization of identical coupled R\"ossler oscillators.}
{\bf (a)} Chaotic R\"ossler oscillators are coupled according to a star graph ($N=8$). This graph has
an EEP with two cells (indicated with colors), shown with its quotient graph.  
{\bf (b)} Under certain conditions, the coupled oscillators 
(each with a three-variable dynamics $\mathbf{x}(t)=(x_1(t), x_2(t), x_3(t))$) 
can exhibit cluster synchronization commensurate with the EEP: 
spoke nodes (blue), centre node (red).
{\bf (c)-(d)} Analogously to linear consensus (Fig.~\ref{fig:schematic_EEP}), given an initial condition 
consistent with the EEP, the dynamics of the nodes 
within each cell remain identical. The solid lines are the full dynamics $\mathbf{x}(t)$
governed by~\eqref{sync} with initial condition ${\bf x}_0=(H \otimes I_d) {\bf y}_0$; 
the circles are the quotient graph dynamics 
$\mathbf{y}$ governed by~\eqref{eq:sync_quotient}. 
Figures {\bf (c)} and {\bf (d)} correspond to two values of the coupling parameter $\gamma$,
and in both cases the dynamics remains cluster-synchronized within the EEP. 
In {\bf (c)}, with $\gamma=0.3$, the total synchronization of the 
quotient graph is stable, and both $\mathbf{x}$ and $\mathbf{y}$ 
converge to the completely synchronized solution. 
In {\bf (d)}, with $\gamma=0.03$, total synchronization of the quotient graph dynamics
is not linearly stable; hence the system exhibits sustained cluster synchronization.} 

\label{figsync}
\end{figure}

\subsubsection{EEPs and invariance of cluster-synchronized states}

Let a graph with Laplacian $L$ exhibit a non trivial EEP with $c$ cells 
encoded by the indicator matrix $H$ and quotient Laplacian $L^\pi$. 
The dynamics of the cell variables $\mathbf{y} \defeq [{\bf y}_1^T,...,{\bf y}_c^T]^T \in \mathbb{R}^{c \, d}$ 
associated with the quotient graph is then given by
\begin{flalign}
    \label{eq:sync_quotient}
\dot{\bf y}&=\mathbf{F}_c({\bf y}) - \gamma (L^\pi \otimes I_d) {\bf G}_c({\bf y}),
\end{flalign}
where $\mathbf{F}_c(\mathbf{y}), \mathbf{G}_c(\mathbf{y}) \in \mathbb{R}^{c \, d}$ 
are defined analogously to $\mathbf{F}_N, \mathbf{G}_N$ above, and
we have the relations:
\begin{align} 
\label{eq:F_G_commute}
(H\otimes I_d)\mathbf{F}_c(\mathbf{y}) &= \mathbf{F}_N \left ( (H\otimes I_d) \mathbf{y} \right) \\
(H\otimes I_d)\mathbf{G}_c(\mathbf{y}) &= \mathbf{G}_N \left ( (H\otimes I_d)\mathbf{y} \right).
\end{align}

In close parallel to the linear case~\eqref{eq:invariant_cells}, 
we can derive the following result for cluster-synchronized dynamics.
Let us have an initial condition that is identical within the cells of the EEP, i.e., ${\bf x}=(H\otimes I_d){\bf y}$ for some arbitrary ${\bf y}\in \mathbb{R}^{cd}$ at $t=0$.
Then the nodes within cells of the EEP remain identical for all time $t \geq 0$, 
and their dynamics can be described by the dynamics of the quotient graph:
\begin{align}
\label{eq:MSF_cells}
\dot{{\bf x}}&=(H\otimes I_d)\dot{{\bf y}} \\
\text{ where }   \quad  
\dot{\mathbf{y}} & = \mathbf{F}_c({\bf y}) - \gamma(L^\pi \otimes I_d){\bf G}_c({\bf y}).\nonumber
\end{align}
This result follows from:
\begin{flalign*}
    \dot{\mathbf{x}} &= (H \otimes I_d) \dot{\mathbf{y}} = 
    (H \otimes I_d) \left[\mathbf{F}_c({\bf y}) - \gamma(L^\pi \otimes I_d){\bf G}_c({\bf y}) \right]\\
    &=  \mathbf{F}_N \left ( (H\otimes I_d) \mathbf{y} \right) - \gamma(HL^\pi \otimes I_d ) {\bf G}_c({\bf y})\\
    &= \mathbf{F}_N \left ( (H\otimes I_d) \mathbf{y} \right) - \gamma(LH \otimes I_d ) {\bf G}_c({\bf y})\\
    &= \mathbf{F}_N \left ( (H\otimes I_d) \mathbf{y} \right) - \gamma(L \otimes I_d)(H\otimes I_d) {\bf G}_c({\bf y}) \\
    &=\mathbf{F}_N \left ( (H\otimes I_d) \mathbf{y} \right) - \gamma(L \otimes I_d){\bf G}_N((H\otimes I_d){\bf y}) \quad \square
\end{flalign*}
Here we have made use of the standard identity $(A\otimes B)(C\otimes D) = (AC)\otimes(BD)$. \\

\noindent \textbf{\emph{Example [Coupled R\"ossler oscillators]}}:
\label{ex1sync}
Consider a network of $N=8$ oscillators where 
each node has a three dimensional dynamics ($d=3$) 
given by the chaotic R{\"o}ssler system~\cite{Roessler1976}
\begin{equation}
\label{rosseqn}
{\bf F}(\mathbf{x})= \mathbf{F}([x_1,x_2,x_3]^T) =
\begin{bmatrix}
-x_2 - x_3 \\
x_1 + ax_3 \\
b+x_3(x_1-c)
\end{bmatrix}
\end{equation}
with parameters $a=b=0.2$ and $c=7$. The
oscillators are coupled through the variable $x_1$ according to the linear function:
$$
\mathbf{G}(\mathbf{x}) = 
\begin{bmatrix}
1 & 0 & 0 \\
0 & 0 & 0 \\
0 & 0 & 0
\end{bmatrix} \mathbf{x}.
$$
The topology of interconnection is a star graph, which
has an EEP with two cells ($c=2$): one cell comprises the central node, 
the other cell contains all other nodes~(Fig.~\ref{figsync}).

Let the initial condition be ${\bf x}_0=(H\otimes I_d){\bf y}_0$. Then the variables of the nodes within each cell remain identical at all times, i.e.,  the dynamics stays cluster-synchronized (Fig.~\ref{figsync}).
For $\gamma = 0.3$, this poly-synchronous state (while remaining cluster-synchronized at all times) evolves towards the globally synchronized state (Fig.~\ref{figsync}c). 
In contrast, for $\gamma = 0.03$, the cluster synchronization of the cells does not converge towards global synchrony, since the completely synchronized state of the quotient graph dynamics is no longer (linearly) stable (Fig.~\ref{figsync}d and Sec.~\ref{sec:MSF_stability}).

\subsubsection{EEPs and cell-averaged synchronization dynamics}
Although the invariance of cluster-synchronized EEP states 
carries over to the nonlinear MSF setting,
the second finding of the linear analysis, namely that the dynamics of cell averages is 
described by the quotient graph dynamics~\eqref{eq:sync_quotient}, does not hold in general.
Indeed, after some algebraic manipulations it is easy to see that:
\begin{flalign*}
    &(H^+\otimes I_d) \dot{\mathbf{x}} = (H^+\otimes I_d) [\mathbf{F}_N({\bf x}) - \gamma (L \otimes I_d) {\bf G}_N({\bf x})]\\
   &=(H^+\otimes I_d)\mathbf{F}_N({\bf x}) - \gamma (L^\pi \otimes I_d)(H^+ \otimes I_d) \mathbf{G}_N({\bf x}).
\end{flalign*}
Due to their nonlinearity, in general $\mathbf{F}$ and $\mathbf{G}$ do not commute with the linear cell-averaging operation:
\begin{align*}
(H^+\otimes I_d)\mathbf{F}_N(\mathbf{x}) & \neq \mathbf{F}_c((H^+\otimes I_d)\mathbf{x}) \\
(H^+\otimes I_d)\mathbf{G}_N(\mathbf{x}) & \neq \mathbf{G}_c((H^+\otimes I_d)\mathbf{x}).
\end{align*}
 Hence, unlike the linear case, the cell-averaged dynamics is not strictly equivalent to the synchronization 
 dynamics governed by the Laplacian of the quotient graph.

However, an approximate equivalence is obtained if we consider an $\boldsymbol{\epsilon}$
perturbation around a cluster-synchronized state~\eqref{eq:cluster_state}.
To first order we then have, 
\begin{align} \nonumber
    \mathbf{F}_N \left( \mathbf{x}_s + \boldsymbol{\epsilon} \right ) 
    & \approx \mathbf{F}_N((H \otimes I_d)\mathbf{y}_s) + D \mathbf{F}_N(\mathbf{x}_s) \,\boldsymbol{\epsilon}, \\
        \mathbf{G}_N \left( \mathbf{x}_s + \boldsymbol{\epsilon} \right ) 
    & \approx \mathbf{G}_N((H \otimes I_d)\mathbf{y}_s) + D \mathbf{G}_N(\mathbf{x}_s) \, \boldsymbol{\epsilon},  \nonumber
\end{align}
where $D\mathbf{F}_N(\mathbf{x})$ and $D\mathbf{G}_N(\mathbf{x})$ denote the Jacobians of $\mathbf{F}_N$ and $\mathbf{G}_N$ for state $\mathbf{x}$.

\begin{figure}[t!]
\centering
\includegraphics[width=.45 \textwidth]{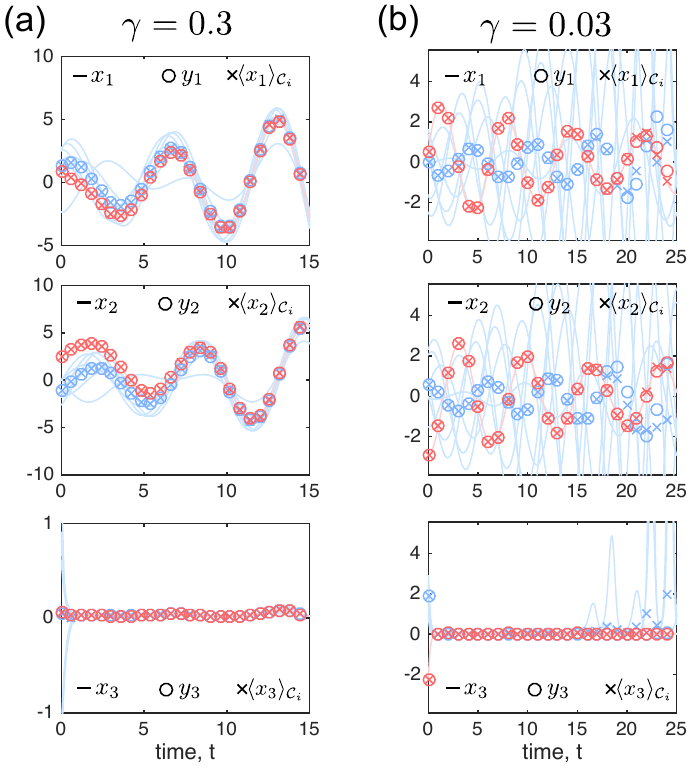}
\caption{\textbf{Cell-averaged dynamics and synchronization of identical R\"ossler coupled oscillators.}
The numerics in this Figure follow Figure~\ref{figsync}, but here we focus on the cell-averaged dynamics of the same system of coupled R\"ossler oscillators and its relationship with the dynamics of the quotient graph. 
{\bf (a)} For $\gamma =0.3$ and an initial condition close to the synchronization manifold, the quotient dynamics (circles) evolves closely to the cell-averaged full system dynamics (crosses), and both dynamics converge to the totally synchronized solution. Note the individual time courses of each of the eight oscillators converging also to this solution.
{\bf (b)} For $\gamma =0.03$, however, the cell-averaged (crosses) and quotient dynamics (circles) diverge, as is clearly visible for large times.
}
\label{figsync2}
\end{figure}

This result implies that if the cluster-synchronized state is stable, the averaging operator will approximately commute with both $\mathbf{F}_N$ and $\mathbf{G}_N$ when the state is close to the cluster-synchronized state. As a consequence, an appropriately chosen initial condition of the average cell dynamics will remain close (or converge) to the quotient dynamics, as seen in Fig.~\ref{figsync2}a. 
On the other hand, the interplay of the Jacobians of $\mathbf{G}$ and $\mathbf{F}$ and the graph structure encoded by $L$ and $L^\pi$ can render the initial perturbation unstable, and the state will exponentially diverge. In that case, the quotient dynamics will not be a good model for the cell-averaged dynamics, as seen in 
Fig.~\ref{figsync2}b.  
We explore these points through the MSF formalism in Section~\ref{sec:MSF_stability}, where we consider the stability of the cluster-synchronized state (including the globally synchronized state).

\subsubsection{Stability of EEP cluster-synchronized states through the MSF formalism}
\label{sec:MSF_stability}

The sections above lead naturally to consider the stability of EEP cluster synchronization.
Following Pecora et al \cite{Pecora2014,Sorrentino2016}, the linearized stability around any cluster 
synchronized state can be evaluated using the MSF framework via the variational expression:
\begin{small}
\begin{flalign}\label{eq:cluster_sync_stab_org}
\delta \dot{\mathbf{x}}(t)& =  \left[ \sum_{i=1}^c \left(E^{(i)} \otimes D\mathbf{F}(\mathbf{s}_i) \right)  - \gamma 
\left(LE^{(i)} \otimes D\mathbf{G} (\mathbf{s}_i) \right) \right] \delta{\bf x}(t),
\end{flalign} 
\end{small}where $\mathbf{s}_i \in \mathbb R^d$ is the (consistent) state of every node in the $i$th cluster, as defined
in~\eqref{eq:cluster_state},
and $E^{(i)}$ are identity matrices consigned to each cluster:
\begin{align}
E^{(i)} & \defeq \text{diag}(\mathbf{h}_i),
\end{align}  
as given by the cell indicator vectors~\eqref{eq:H_cols}.
Using computational group theory, Pecora \textit{et al.} block-diagonalize the above expression 
to assess the stability of any cluster-synchronized state.

As an alternative to symmetry-based arguments, the MSF variational analysis may also be understood
using EEPs and their associated indicator matrices.  Here we use the fact that eigenvectors and eigenvalues are shared between the Laplacians of the original and quotient graphs~\cite{OClery2013}.
Let us denote the $c$ eigenvectors of the quotient Laplacian $L^\pi$ by ${V^\pi = [\mathbf{v}^\pi_1, \ldots, \mathbf{v}^\pi_c]}$ with eigenvalues $\Lambda^\pi = \mathrm{diag}(\lambda_i^\pi)$ such that $L^\pi V^\pi = V^\pi \Lambda^\pi$. 
The properties of the EEP~\cite{OClery2013} ensure that a subset of the eigenvectors of the full Laplacian $L$ are directly related to the eigenvectors of $L^\pi$: 
\begin{align}
\label{eq:Vs_def}
V_\text{s} &= HV^\pi \in \mathbb{R}^{N\times c}.
\end{align} 
These are the eigenvectors that define the cluster synchronization manifold commensurate with the EEP. 
The eigenvectors orthogonal (transversal) to the cluster-synchronized manifold
are denoted by $V_\perp \in \mathbb{R}^{N\times (N-c)}$. These are the eigenmodes that drive the system out of a cluster-synchronized state, and therefore we want these modes to be damped.
An orthogonal matrix of eigenvectors of $L$ that diagonalizes the Laplacian:
\begin{align}
V^TLV & = \Lambda  \quad \text{where} \quad \Lambda \defeq \text{diag}(\lambda_i).
\end{align}
is thus given by 
\begin{align}
V=[V_\text{s}, V_\perp] = [H V^\pi, V_\perp ].
\end{align}
Hence the first $c$ columns correspond to eigenvectors of $L$ (with eigenvalues $\lambda_i =\lambda_i^\pi, i=1,\dots,c$) that can be mapped to $L^\pi$, and the second block of $(N-c)$ columns 
corresponds to the transversal manifold.

Using $V$ to diagonalize $L$ via the coordinate transformation $\delta {\boldsymbol{\chi}}=( V^T \otimes I_n )\delta {\bf x}$ leads to:
\begin{small}
\begin{align}
\delta  \dot{\boldsymbol{\chi}} (t)  & =  ( V^T \otimes I_n ) \Big[ \sum_{i=1}^c E^{(i)} \otimes D\mathbf{F}(\mathbf{s}_i) \nonumber \\
&  \qquad - \gamma \sum_{i=1}^c LE^{(i)} \otimes D\mathbf{G} (\mathbf{s}_i) \Big] ( V \otimes I_n )\delta\boldsymbol{\chi}(t)   \\
& =  \Big[ \sum_{i=1}^c V^T E^{(i)} V \otimes D\mathbf{F}(\mathbf{s}_i)  \nonumber \\
& \qquad - \gamma \sum_{i=1}^c V^T LE^{(i)} V \otimes D\mathbf{G} (\mathbf{s}_i) \Big]\delta\boldsymbol{\chi}(t) \\
\label{eqn:stability}
&=\left[ \sum_{i=1}^c \left(Q^{(i)} \otimes D\mathbf{F}(\mathbf{s}_i) \right)
 - \gamma \left(\Lambda Q^{(i)} \otimes D\mathbf{G} (\mathbf{s}_i) \right) \right] \delta\boldsymbol{\chi}(t), 
\end{align}
\end{small}
where we have 
\begin{align}
V^TLE^{(i)}V & = \Lambda \left(V^TE^{(i)}V \right) \defeqr \Lambda  \, Q^{(i)}.
\end{align}

The structure of the matrices $Q^{(i)}$ means that the modes in the cluster synchronization manifold are effectively decoupled from the modes transversal to it.
To see this, note that  from  $V_\perp^T V_\text{s} =0$ and~\eqref{eq:Vs_def} it follows 
that the transversal eigenvectors $V_\perp$ lie in the orthogonal subspace to $H$: 
$H^T V_\perp = 0$.
(This also means that every transversal mode is mean-free within each cell: $H^+ V_\perp = 0$.)
Therefore, we have the following effective decoupling between the cluster-synchronized and transversal modes:
\begin{equation}
V_\perp^TE^{(i)}V_\text{s} = V_\perp^TE^{(i)}HV^\pi= V_\perp^T[0,\ldots, \mathbf{h}_i,0,\ldots]V^\pi =0, \nonumber
\end{equation}
leading to 
\begin{align*}
Q^{(i)}&=V^T E^{(i)} V =  
 \begin{bmatrix}
Q_s^{(i)} & 0_{c\times (N-c)} \\
  0_{(N-c)\times c} & Q_\perp^{(i)}
 \end{bmatrix}.  
\end{align*}
By examining this matrix, we can obtain information about the (local) stability of the cluster-synchronized state 
(see Ref.~\cite{Sorrentino2016} for a related discussion).
In order to check the linear stability of the cluster-synchronized manifold,
it is enough to check that all the transversal modes are damped. 
Yet such damping of the transversal modes alone does not specify the behavior \emph{within} the cluster-synchronized manifold, or indeed the convergence towards any of the different cluster-synchronized states within it.  
Further damping \emph{within} the cluster-synchronized manifold would lead
the dynamics to converge to an even lower-dimensional manifold, i.e., towards a particular subset of the cluster-synchronized states.
Stated differently, some of the cells in a cluster-synchronized state could merge, leading to another state with fewer cells.
If damping within the manifold is present, it can lead to convergence towards the completely synchronized state, akin to the numerics in Fig.~\ref{figsync}c  (and in contrast to the numerics in Fig.~\ref{figsync}d where such convergence within the manifold is not observed).

\subsection{EEP cluster synchronization in Kuramoto networks}\label{sec:kuramoto_pos}
The MSF framework provides a powerful tool for the analysis of nonlinear systems with 
diffusive couplings, yet there are important classes of systems that do not lend 
themselves naturally to this formulation. Examples include systems with sinusoidal coupling 
between oscillators, as in models of power systems~\cite{Doerfler2012} or 
the classic Kuramoto model of coupled oscillators~\cite{Kuramoto1975,Kuramoto2012}.
The use of EEPs can nevertheless afford us insight into cluster synchronization in these cases, too. 

\begin{figure*}
\includegraphics[]{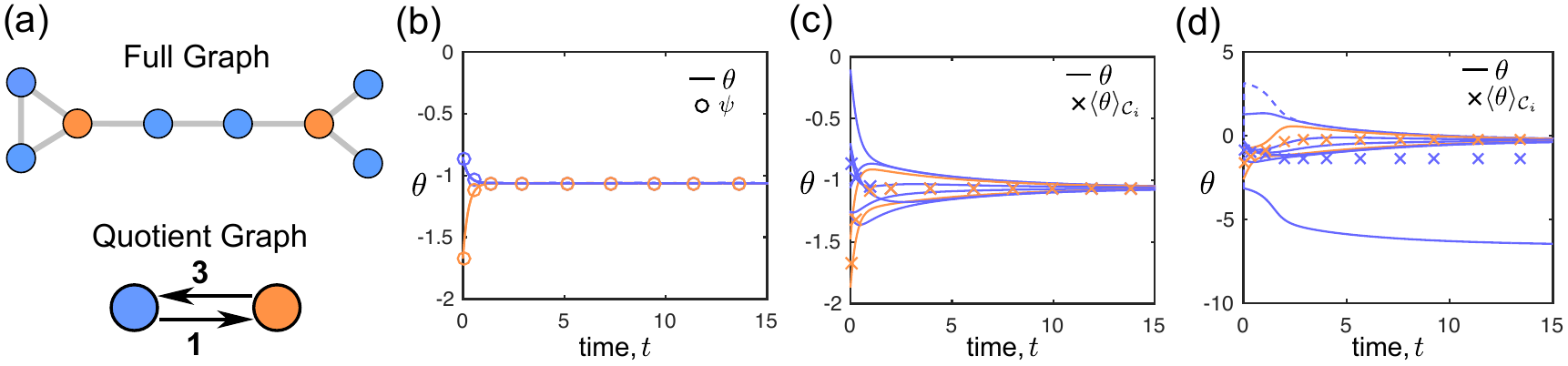}
\caption{\textbf{EEP clustered dynamics on a network of Kuramoto oscillators with identical intrinsic frequencies.} \textbf{(a)} Kuramoto oscillators ($N=8$) coupled through the graph shown, which has an EEP indicated by the color of the nodes. This partition is not an orbit partition, i.e., it is not directly induced by any symmetry group. The associated quotient graph is shown below. 
    \textbf{(b)} If the initial condition is constant within each cell, the dynamics of the full system (line) is exactly equivalent to the lower dimensional Kuramoto dynamics on the quotient graph (circles). 
    \textbf{(c)} 
    Provided the initial condition for ${\boldsymbol{\theta}}$ is close to the cell-averaged state and within an open semi-circle, the linear cell- averaged dynamics (crosses) is closely aligned with the quotient graph dynamics shown in (b).
    \textbf{(d)} If the initial condition is spread further on the circle, the linear cell averaging is no longer aligned with the quotient graph.
    }
\label{fig:kuramoto}
\end{figure*}

We consider the Kuramoto model with $N$ oscillators
\begin{equation}
    \frac{d \theta_i}{d \tau} = \omega_i + \lambda \sum_{j=1}^N A_{ij}\sin(\theta_j - \theta_i),
\end{equation}
where $\theta_i$ and $\omega_i$ describe the phase and intrinsic frequency of each oscillator,  
respectively, $\lambda$ is the coupling parameter, and $A_{ij}$ is the adjacency matrix encoding the network connectivity.

To simplify our notation below, let us renormalize time $t = \lambda \tau$.  
The dynamics of a network of Kuramoto oscillators coupled through a graph with Laplacian $L = BB^T$, where $B$ is the incidence matrix of the graph, can then be rewritten 
in vector-matrix notation as~\cite{Jadbabaie2004}: 
\begin{align}  
\label{eq:kuramoto_matrix}  
\dot{\boldsymbol{\theta}} 
&= \frac{1}{\lambda} \boldsymbol{\omega}  - B\sin(B^T{\boldsymbol{\theta}}) \\
      &= \frac{1}{\lambda}  \boldsymbol{\omega} - B\mathcal W(B^T{\boldsymbol{\theta}}) B^T \, {\boldsymbol{\theta}}  \nonumber \\
   & =\frac{1}{\lambda}   \boldsymbol{\omega}  -  \LW{\mathcal{W}} {\boldsymbol{\theta}} \, \boldsymbol{\theta},
\end{align}
where $\boldsymbol{\theta}$ and $\boldsymbol{\omega}$ are $N$-dimensional vectors, and we have defined 
$\mathcal W(\mathbf{x}) \defeq \text{diag}( \mathrm{sinc}(\mathbf{x})) = \mathrm{diag}(\sin(x_i)/x_i).$

This rewriting emphasizes the close relation of the Kuramoto model to Laplacian dynamics. Not only
does the linearization for small phase differences lead to the standard linear Laplacian dynamics,
but the final equality underscores the fact that the full Kuramoto model 
may still be understood in terms of a weighted Laplacian dynamics
with time-varying edge weights~\cite{Jadbabaie2004}:
\begin{align}
\label{eq:weighted_Laplacian}
\LW{\mathcal{W}} {\boldsymbol{\theta}} \defeq B\mathcal W(B^T{\boldsymbol{\theta}}) B^T.
\end{align}
It is therefore not surprising that EEPs give useful insights into invariant dynamics of Kuramoto networks.

\subsubsection{Case I: equal intrinsic frequencies}
\label{sec:Kuramoto_equal}
Let us consider first the case where all intrinsic frequencies are identical: $\omega_i = \omega,\; \forall i$. 
In this case, we may assume $\boldsymbol{\omega} =0$ without loss of generality, as this is equivalent to grounding the system or defining the phases with reference to a rotating frame~\cite{Jadbabaie2004}. 
The resulting system
\begin{equation}\label{eq:kuramoto_large}
    \dot {\boldsymbol{\theta}} = -B\sin(B^T{\boldsymbol{\theta}}) 
    = - \LW{\mathcal W}{\boldsymbol{\theta}} \, {\boldsymbol{\theta}}
\end{equation}
is well known to converge~\cite{Jadbabaie2004,Doerfler2014} to the totally synchronized state with identical phases.

Let the graph with Laplacian $L=B B^T$ be endowed with an EEP with partition matrix $H$.
We can then define the following Kuramoto dynamics taking place on the quotient graph of the EEP:
\begin{equation}\label{eq:kuramoto_small}
    \dot {\boldsymbol{\psi}} =-H^+B \, \sin(B^TH{\boldsymbol{\psi}}) = 
    -H^+ \LW{\mathcal W}{H \boldsymbol{\psi}} \, H {\boldsymbol{\psi}}
\end{equation}
where ${\boldsymbol{\psi}}$ is the $c$-dimensional vector containing the phases associated with the quotient graph, and we use the definition~\eqref{eq:Lpi} to factorize the quotient Laplacian appropriately: 
\begin{equation}
L^\pi = H^+ L H = (H^+ B) (B^T H).
\end{equation}

As for the linear case~\eqref{eq:invariant_cells}, we wish to show that the cell dynamics on the quotient
graph~\eqref{eq:kuramoto_small} describes an invariant dynamics of cluster-synchronized states in the 
full model~\eqref{eq:kuramoto_large}. In other words, we need to show that 
\begin{flalign}
\label{eq:psi_fulldyn}
H\dot {\boldsymbol{\psi}} = -B\sin(B^T H{\boldsymbol{\psi}}), 
\end{flalign}
i.e., $H{\boldsymbol{\psi}}$ is invariant under the full dynamics.

To establish this, we use the following fact:

\textit{ A given EEP for a network remains an EEP  if all edge weights 
between two distinct cells are multiplied by a factor that depends only on the two cells.}

This fact is a direct consequence of the definition of an EEP, since such scaling changes the
out-degree patterns of all nodes within a cell consistently.
\\

\refstepcounter{remarkcounter}
\noindent \textbf{\emph{Remark \theremarkcounter}} \textbf{\emph{[EEPs and structured
weights]:}}
\textit{
A particular case of such a rescaling that will be useful below can be represented algebraically as follows.
Consider a graph with Laplacian $L= BB^T$ and an EEP with indicator matrix $H$. 
Let the edge weights be scaled consistently across cells (in the above sense) leading to the modified Laplacian:
\begin{equation}
\LW {w} {H\boldsymbol{\xi}}  = B \, \mathrm{diag}(w(B^TH\boldsymbol{\xi})) \, B^T,
\end{equation} 
where $\boldsymbol{\xi}\in \mathbb R^c$ is a cell vector, and $w(x) = w(-x)$ is a symmetric function applied element-wise. 
Since the EEP remains unchanged under this rescaling, it follows from \eqref{eq:PH_commutes} that the projection operator associated with $H$ also commutes with the modified Laplacian: 
\begin{equation}
\label{eq:PH_commutes_weighted}
P_H L = L P_H  \implies
P_H  \LW {w} {H\boldsymbol{\xi}} = \LW {w} {H\boldsymbol{\xi}} P_H. 
\end{equation} }

We now use~\eqref{eq:PH_commutes_weighted} to show that EEP cluster-synchronized states are 
invariant under Kuramoto dynamics.
To see this, left multiply~\eqref{eq:kuramoto_small} with $H$:
\begin{align*}
H\dot {\boldsymbol{\psi}} &= -HH^+B\sin(B^TH{\boldsymbol{\psi}})  \\
&=     -P_H \LW{\mathcal W}{H \boldsymbol{\psi}} \, H {\boldsymbol{\psi}} 
= - \LW{\mathcal W}{H \boldsymbol{\psi}} P_H \, H {\boldsymbol{\psi}} \\
& = - \LW{\mathcal W}{H \boldsymbol{\psi}} \, H {\boldsymbol{\psi}}= -B\sin(B^T H{\boldsymbol{\psi}}), \nonumber
\end{align*}
where $P_H H = H$ follows from the definition of the projection operator. Note that $\mathcal{W}(x) = \mathrm{sinc}(x)$ is symmetric.~$\square$

The proof shows that the full Kuramoto model follows the quotient dynamics~\eqref{eq:kuramoto_small} for all times, if it ever synchronizes to a particular EEP. 
We illustrate this behavior in Fig. \ref{fig:kuramoto}b, where we use a network topology (Fig.~\ref{fig:kuramoto}a) inspired by a construction outlined by Chan and Godsil~\cite{Chan1997,Kudose2009} highlighting the difference between orbit partitions (generated from symmetry groups) and equitable partitions. 

As shown in Fig.~\ref{fig:kuramoto}c, the cell averages are also well described by the quotient dynamics 
provided the initial condition is not too far away from the EEP-averaged state. If the phases of the initial condition are outside the open semicircle (as in Fig.~\ref{fig:kuramoto}d), a naive linear averaging does not fully capture the convergence on the torus.

\begin{figure}[tb!]
\center
\includegraphics[width=.45\textwidth]{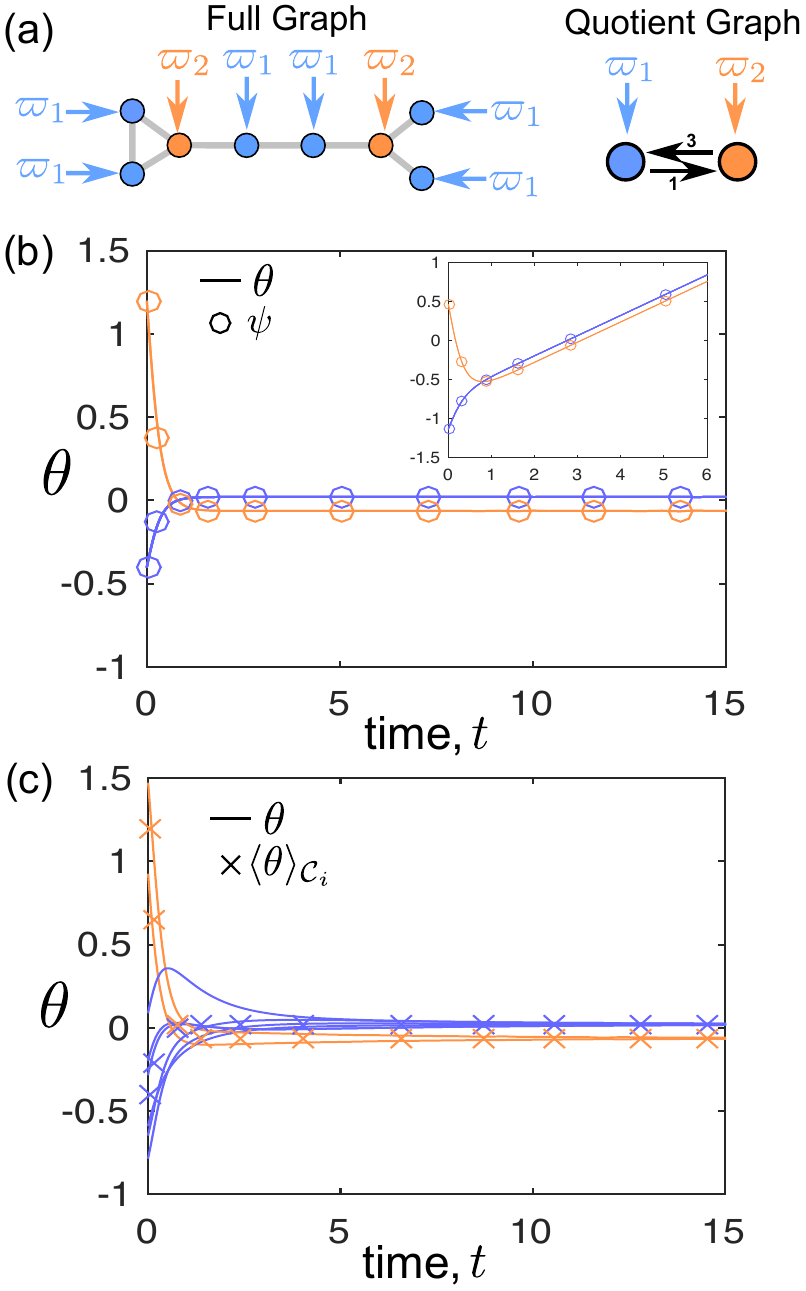}
\caption{\textbf{Cluster synchronization in Kuramoto networks with EEP-commensurate 
intrinsic frequencies.} 
(a)  Kuramoto dynamics~\eqref{eq:kuramoto_large_with_aligned_inputs} over the same network as in Figure \ref{fig:kuramoto}a, but this time with non-identical intrinsic frequencies, yet aligned with the EEP. 
    \textbf{(b)} If the initial condition is constant within cells, the dynamics of the full system (lines) is identical to the dynamics of the quotient graph (circles), and the system eventually settles to a cluster-synchronized state.
    Inset: the same dynamics without subtracting the (time-dependent) mean phase.  
    \textbf{(c)} If the initial phases are within an open semicircle and close to the cell-averages, then the quotient dynamics is a good descriptor for the dynamics for all times. 
    }
\label{fig:kuramoto_diff_freq}
\end{figure}

\subsubsection{Case II: non-equal intrinsic frequencies commensurate with an EEP}
The analysis for the Kuramoto model with equal frequencies does not apply in general 
to a network of oscillators with non-equal intrinsic frequencies.
However, similar results hold when the oscillators within each cell have the same frequency.
In particular, consider the Kuramoto system~\eqref{eq:kuramoto_matrix} with EEP-commensurate
frequencies:
\begin{equation}\label{eq:kuramoto_large_with_aligned_inputs}
    \dot {\boldsymbol{\theta}} = \frac{1}{\lambda} H\boldsymbol{\varpi} -B\sin(B^T{\boldsymbol{\theta}}),
\end{equation}
where $\boldsymbol{\varpi}$
is a $c$-dimensional vector containing the frequencies of the cells. 

In the case of heterogeneous frequencies, the model can not reach globally identical synchronization, so the  `most synchronous' behavior is the cluster-synchronized state with identical phases within each cell. 
By the arguments in~Section~\ref{sec:Kuramoto_equal}, \textit{mutatis mutandis}, it is easy to see that the cluster-synchronized state
$H \boldsymbol{\psi}$ is invariant under~\eqref{eq:kuramoto_large_with_aligned_inputs} and
governed by the quotient graph:
\begin{equation}
  \dot {\boldsymbol{\psi}} = \frac{1}{\lambda} \boldsymbol{\varpi} -H^+B \, \sin(B^TH{\boldsymbol{\psi}}).
\end{equation}
A numerical illustration of this invariance is given in Figure \ref{fig:kuramoto_diff_freq}.
We note that there is a close analogy here to the scenario of the linear 
consensus system with an input commensurate with the EEP~\eqref{eq:consensus_input}.
Indeed, the intrinsic frequencies of the cells $\boldsymbol{\varpi}$ can be 
interpreted as constant inputs to each of the cells.

We remark that our results for Kuramoto systems here are concerned with the \emph{invariance of solutions} 
and not their stability.  As studied previously~\cite{Jadbabaie2004,Doerfler2014}, 
the stability of the synchronous state depends on the magnitude of the spread of the 
frequencies $\omega_i$ along the edges of the graph relative to the coupling parameter $\lambda$. 
Hence as the coupling $\lambda$ becomes smaller, and the norm of $\boldsymbol{\varpi}/\lambda$ becomes larger, the synchronized (and cluster-synchronized) solutions become unstable. 
\\

\refstepcounter{remarkcounter}
\noindent \textbf{\emph{Remark \theremarkcounter}} \textbf{\emph{[Kuramoto model with a phase offset]:}}
\label{rem_phase_offset} 
\textit{To gain insight into the effect of a phase offset, let us consider the Kuramoto model with equal intrinsic frequencies and a constant phase offset discussed in Ref.~\cite{Nicosia2013}, which can be rewritten as:
\begin{align}
\label{eq:kuramoto_phase}
    \dot {\boldsymbol{\theta}} & = -B\sin(B^T{\boldsymbol{\theta}} + \alpha \mathbf{1}) \\
& = \sin(\alpha)B\cos(B^T{\boldsymbol{\theta}}) -\cos(\alpha) B \sin(B^T{\boldsymbol{\theta}}) .
\end{align}
For $\alpha \to 0$, the first term vanishes, and we recover the standard Kuramoto model~$\eqref{eq:kuramoto_large}$ for which the EEP analysis holds. 
Therefore, if ${\boldsymbol{\theta}}$ is within (close proximity to) the cluster synchronization manifold $H \boldsymbol{\psi}$, by our arguments above, the system will remain in a polysynchronous, clustered state.
As $\alpha$ increases, the magnitude of the Kuramoto coupling parameter ($\cos \alpha)$ decreases, whereas at the same time the magnitude of the spread of the input intrinsic frequencies ($\sin \alpha)$ increases. 
Hence, as $\alpha$ is increased above a threshold, we expect the cluster synchronization manifold to lose stability, as for the case of non-equal frequencies above. This is in line with~\citet{Nicosia2013}, who observed numerically that cluster synchrony is lost above a critical value of $\alpha$.}

\section{Cluster synchronization in networks with positive and negative weights}
\label{sec:negative_wts}
Many mathematical models for real-world networks need to incorporate positive and negative interactions. 
For instance, in social networks, relationships can be friendly or hostile, or they reflect trust or 
distrust between individuals.  
Therefore, the sign of a link is a central concept in social psychology, associated with the emergence of conflict and tension in social systems~\cite{Heider1946,Cartwright1956}, and it has gained popularity recently in the study of online social networks \cite{Leskovec2010} and online cooperation \cite{Szell2010}. 
In biological systems, the sign of an edge is also a key element, in particular when modelling dynamical processes.  For instance, genes can either promote or repress the expression of other genes in genetic regulatory networks~\cite{Davidson2005}, and neurons can excite or inhibit the firing of other neurons in neuronal networks and thereby shape the global dynamics of the system~\cite{Gerstner2014,Schaub2015}.

\subsection{Signed networks and structural balance: the signed external equitable partition}
\subsubsection{The signed Laplacian matrix}
For a network with positive and negative interactions, we can define the \emph{signed Laplacian matrix} of the network as follows \cite{Altafini2013,Kunegis2010}:
\begin{equation}
L_\sigma = D_\text{abs} - A,
\end{equation}
where $D_\text{abs} = \text{diag}(|A|\mathbf{1})$ is the diagonal \textit{absolute} degree matrix, 
and $A$ is again the adjacency matrix (which may hereafter contain both positive and negative weights).
As for the standard Laplacian, it can be shown that the signed Laplacian is positive semidefinite and 
its spectrum contains one zero eigenvalue when the graph is connected and structurally balanced.
To see this, note that the signed Laplacian can be expressed as:
\begin{equation}
    L_\sigma = B_\sigma^{}W_\text{abs} B_\sigma^T,
\end{equation}
where $W_\text{abs} =  \text{diag}(|w_e|)$ is the absolute edge weight matrix and $B_\sigma\in \mathbb R^{N\times E}$ is the signed node-to-edge incidence matrix:
\begin{equation*}
    [B_\sigma]_{ie} = \begin{cases} 1& \text{ if $i$ is the tail of edge $e$}, \\ -\text{sign}(e)& \text{ if $i$ is the head of edge $e$,}\end{cases}
\end{equation*}
Henceforth, we assume $W_\text{abs} = I$ without loss of generality.
By using the signed Laplacian, the construction of an EEP can be extended to signed graphs.
To do so, however, we must first introduce the notion of structurally balanced graph, which will enable us to define the notion of a signed external equitable partition (sEEP). 

\subsubsection{Structurally balanced graphs}
\begin{figure}
\center
\includegraphics{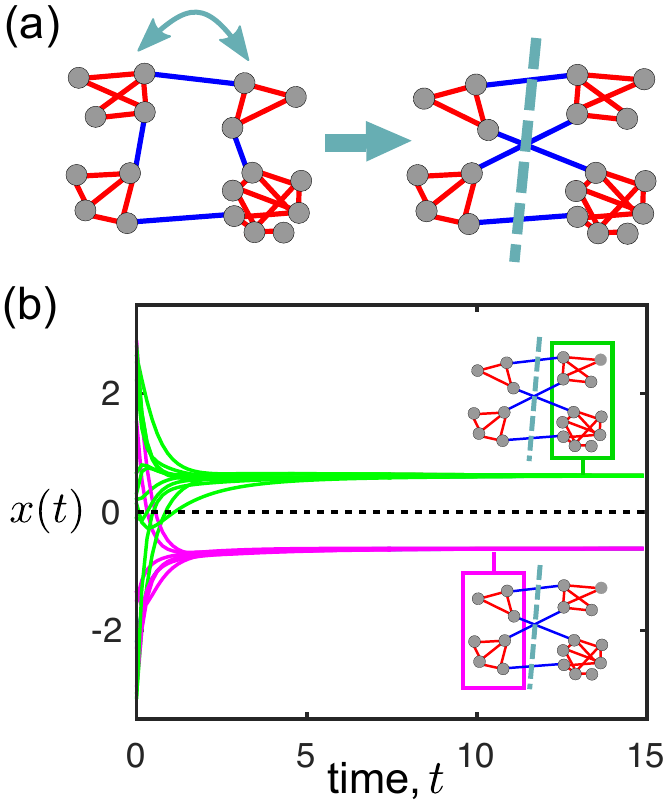}
\caption{\textbf{Structurally balanced graphs and bipolar consensus}. 
    \textbf{(a)} Example of a structurally balanced signed graph (red links positive, blue links negative).  
    Every cycle has an even number of negative links or, equivalently, the graph can divided into two factions given by their polarization $\sigma_i$ (corresponding to the green and magenta groups).  Note that each of the factions has only positive weights inside and only negative weights between them.    
    \textbf{(b)} The signed consensus dynamics~\eqref{eq:signed_consensus} on a structurally balanced graph always leads to a bipolar consensus, in which each node agrees with the nodes within its own faction, but has exactly the opposite sign to any node in the other faction.  }
\label{fig:struct_balanced_graph}
\end{figure}

Following Cartwright and Harary~\cite{Cartwright1956}, a signed graph is defined to be \emph{structurally balanced} if the product of the signs along any closed path in the network is positive.
This definition implies that only `consistent' social relationships are allowed in triangles of three nodes: either all interactions are positive, or there are exactly 2 negative links, which may be interpreted in the sense that ``the enemy of my enemy is my friend''~\cite{Heider1946}.  
Equivalently, a signed network is structurally balanced if it can be split into two factions, where each faction contains only positive interactions internally, while the connections between the two factions are purely antagonistic (see Fig.~\ref{fig:struct_balanced_graph}).
It has been shown that many social networks are close to being structurally balanced~\cite{Facchetti2011}, suggesting that there might be a dynamical process acting on such systems driving them towards structural balance~\cite{Marvel2011,Traag2013}.

The following characterization of a structurally balanced graph based on the signed Laplacian was highlighted by Altafini~\cite{Altafini2013,Altafini2013a}. A network is structurally balanced if there exists a diagonal matrix $\Sigma = \mathrm{diag}(\boldsymbol{\sigma})$, with $\pm 1$ on the diagonal, such that the matrix:
\begin{equation}
\label{eq:positive_switch_equivalent}
    L' = \Sigma L_\sigma \Sigma
\end{equation}
contains only negative elements on the off-diagonal. In other words, the signed Laplacian can be transformed into the standard Laplacian of an associated graph with only positive weights through the similarity transformation defined by $\Sigma$. 
The matrix $\Sigma$ is called switching equivalence, signature similarity, or gauge transformation in the literature~\cite{Altafini2013a}.
Using this characterization, one can efficiently determine whether a network is structurally balanced~\cite{Facchetti2011} and obtain the corresponding switching equivalence matrix $\Sigma$.
Note that it follows trivially that a standard network with only positive weights is always structurally balanced with $\Sigma = I_N$.   Hence $L_\sigma$ is a generalization of the standard Laplacian.

\subsubsection{Signed external equitable partitions}
Using the signed Laplacian, we extend the concept of EEP to structurally balanced signed networks. 
Consider a structurally balanced signed graph with signed Laplacian $L_\sigma$ and denote the Laplacian of the positive switching equivalent graph by $L' =  \Sigma L_\sigma \Sigma$.
Let $H$ denote the indicator matrix of an EEP of $L'$:
\begin{equation}
    L'H = H \LpiS .
\end{equation}
Then there exists a signed indicator matrix $H_\sigma = \Sigma H$ that
defines an invariant subspace for $L_\sigma$: 
\begin{equation} \label{eq:signed_EEP}
L_\sigma H_\sigma = H_\sigma \LpiS,
\end{equation}
which follows from the definition~\eqref{eq:positive_switch_equivalent} and $\Sigma^2 = I_N$.

\begin{figure*}
\center
\includegraphics[width=0.9 \textwidth]{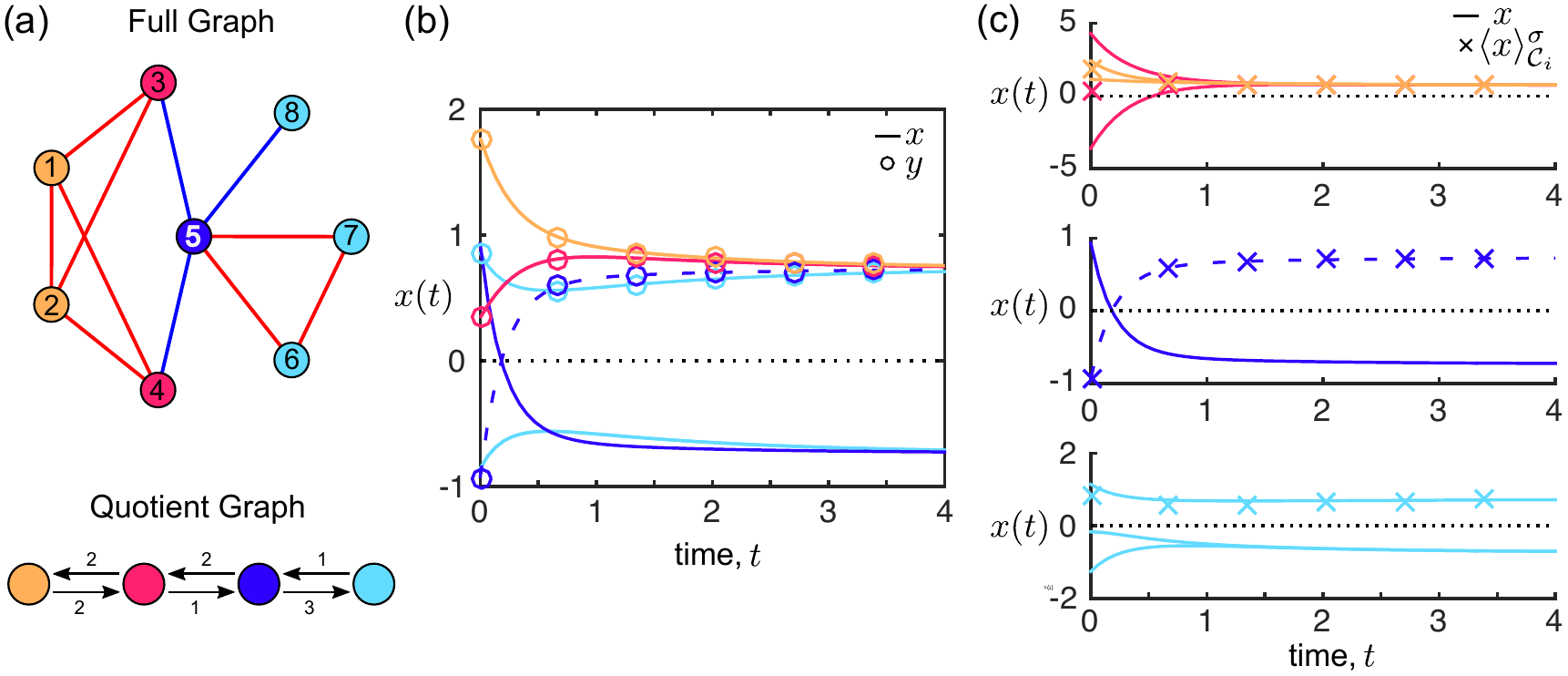}
\caption{\textbf{Signed external equitable partitions and bipolar clustered consensus dynamics.} 
\textbf{(a)} A signed graph (red links are positive, blue links are negative) with a sEEP with four cells (indicated by colors). 
Note how one of the cells (cyan) contains nodes with different polarizations  and another cell (blue) is of negative polarization. 
The associated quotient graph is also shown (bottom).
\textbf{(b)} Similar to the standard graphs with positive weights, if the full dynamics is given by $H_\sigma y$ at any time, then the full dynamics will be exactly determined by the quotient dynamics, but potentially having the opposite sign, like the node in the blue cell with negative polarization
 (whose negative trajectory is shown as a dashed line to make this apparent). 
\textbf{(c)} The sign-adjusted cell averages $\langle x \rangle_{\mathcal{C}_i}^\sigma = H_\sigma^+x$ (crosses) are also determined by the quotient dynamics. 
Trajectories from a random initial condition are shown as solid lines.
Note how the trajectory of the blue node has the opposite sign to its sign-adjusted cell average, and in the cyan cell two of the three nodes (with negative polarization) converge to the sign-flipped value.  }
\label{fig:signed_EEP_consensus}
\end{figure*}

We define the partition $\pi_\sigma$ with indicator matrix $H_\sigma  = \Sigma H$ 
as the \emph{signed external equitable partition} (sEEP), and its associated quotient graph is given by:
\begin{equation}
    \LpiS = H_\sigma^+ L_\sigma H_\sigma = H^+ L' H.
\end{equation}
Therefore cells in a sEEP contain nodes with the same out-degree pattern in absolute value.
An illustration of a sEEP and associated quotient graph is shown in Fig.~\ref{fig:signed_EEP_consensus}.
Note that the quotient graph only has positive weights.

\subsection{Dynamics and signed external equitable partitions} 
The definition of a sEEP provides us with an appropriate tool for the analysis of cluster synchronization in structurally balanced signed networks, as we now show. The results in this section parallel those obtained for unsigned graphs, hence we concentrate on the distinctive features of
clustered dynamics in signed networks.
\subsubsection{The linear case: bipolar cluster synchronization in signed consensus dynamics}
A remarkable feature of structurally balanced networks is that the linear signed consensus dynamics~\cite{Altafini2013}:
\begin{equation}
\label{eq:signed_consensus}
    \dot{\mathbf{x}} = -L_\sigma \mathbf{x},
\end{equation}
converges to a polarized state, in which the nodes are divided into two sets with final values that are equal in magnitude but opposite in sign (Fig.~\ref{fig:struct_balanced_graph}). 
Stated differently, the eigenvector of $L_\sigma$ associated with the zero eigenvalue has the form 
$\boldsymbol{\sigma} = [\sigma_1,\ldots,\sigma_N]^T$, where $\sigma_i\in \{-1,+1\}, \, \forall i$.
As shown in Ref.~\cite{Altafini2013}, this implies that the system dynamics \eqref{eq:signed_consensus} converges to the final state 
\begin{equation}\label{eq:final_state_signed_consensus}
    \lim_{t\rightarrow \infty} \mathbf{x}(t) = 
    \dfrac{\boldsymbol{\sigma}^T \mathbf{x}_0}{N} \boldsymbol{\sigma},
\end{equation}
and the sign pattern of the eigenvector $\mathbf{\sigma}$ corresponds precisely to the switching equivalence transformation, i.e., $\Sigma = \mathrm{diag }(\boldsymbol{\sigma})$.
In the following, we will refer to $\Sigma_{ii} = \sigma_i$ as the polarization of node $i$.
Note that the the vector $\boldsymbol{\sigma}$ is only defined up to an arbitrary sign, so only the relative polarization of the nodes is relevant.

We now extend the analysis to networks endowed with a sEEP.  
The presence of a sEEP~\eqref{eq:signed_EEP} has similar dynamical implications to the presence of an EEP in the case of a positive graph. 
The following statements can be proved analogously to the standard (unsigned) consensus case.

First, sEEP cluster-synchronized states $H_\sigma \mathbf{y}(t)$ are invariant under the full linear dynamics~\eqref{eq:signed_consensus}. 
Hence, given an initial condition $\mathbf{x}=H_\sigma \mathbf{y}$ consistent with a sEEP, $\mathbf{x}(t)$ remains in the sEEP state $\mathbf{x}(t) = H_\sigma \mathbf{y}(t)$ for all times, and the dynamics is governed by the quotient graph: $\dot{\mathbf{y}}  = - \LpiS \mathbf{y}$  (see Figure~\ref{fig:signed_EEP_consensus}b).

In contrast to standard unsigned graphs, the variable of every node within a cell of the cluster-synchronized state will have the same magnitude, but its sign may be inverted depending on its polarization $\sigma_i$.
Therefore, in signed networks each cell maybe itself divided into two factions whose values are of equal magnitude (as given by the quotient dynamics), but of \emph{opposite sign}, as illustrated in 
Figure~\ref{fig:signed_EEP_consensus} (see how node $8$ has the opposite sign to nodes $6,7$ all in the cyan cell).
We use the term \emph{bipolar cluster synchronization}  to account for this phenomenon in signed networks.

Second, the signed cell-averaged dynamics $\langle \mathbf{x} \rangle_{\mathcal{C}_i}^\sigma =H_\sigma^+ \mathbf{x}$ is determined by the dynamics of the quotient graph 
(Fig.~\ref{fig:signed_EEP_consensus}c).
The signed consensus dynamics approaches the bipolar consensus~\eqref{eq:final_state_signed_consensus},
and the final sign of each node variable is determined by $\sigma^Tx_0$, as seen in 
Fig.~\ref{fig:signed_EEP_consensus}c for nodes $1-4$ and $8$ (positively polarized) and 
nodes $5,6,7$ (negatively polarized). 

Third, a system with inputs aligned with the sEEP
\begin{equation}
\label{eq:consensus_input_signed}
    \dot{\mathbf{x}} = -L_\sigma \mathbf{x} + H_\sigma \mathbf{u} ,
\end{equation}
exhibits a bipolar cluster-synchronized state.
To see this, consider the component of the state orthogonal to the bipolar cluster synchronization manifold:
\begin{equation}
\boldsymbol{\delta}(t) \defeq  (I - H_\sigma H_\sigma^+) \, \mathbf{x}(t),
\end{equation} 
which has a dynamics $\dot{\boldsymbol{\delta}} = -L_\sigma \boldsymbol{\delta}$, and the orthogonal component decays asymptotically to 
\begin{equation}
    \lim_{t\rightarrow \infty} \boldsymbol{\delta}(t) = 
    \dfrac{\boldsymbol{\sigma}^T \boldsymbol{\delta}_0}{N} \boldsymbol{\sigma} = \mathbf{0},
\end{equation}
since $\boldsymbol{\sigma}^T \boldsymbol{\delta}_0 =
 \boldsymbol{\sigma}^T  (I - H_\sigma H_\sigma^+) \, \mathbf{x}_0 = 
  (\boldsymbol{\sigma}^T  - \boldsymbol{\sigma}^T) \, \mathbf{x}_0 = 0.$
Hence the system converges to the sEEP manifold.

\subsubsection{Bipolar cluster synchronization for nonlinear dynamics with Laplacian couplings}
All the results obtained in Section~\ref{sec:MSFetc} apply
to nonlinear dynamics on signed networks of the form:
\begin{equation}
    \dot{\mathbf{x}}_i =\mathbf{F}(\mathbf{x}_i) -  \gamma \sum_j [L_\sigma]_{ij} \mathbf{G}(\mathbf{x}_j),
\end{equation}
but now with the additional feature that the dynamics can support a bipolar cluster synchronization based on a sEEP. We do not discuss this case in detail again, instead we illustrate these findings for signed Kuramoto networks.

\subsubsection{Bipolar cluster synchronization in signed Kuramoto networks}
\begin{figure}
\center
\includegraphics[width=.45 \textwidth]{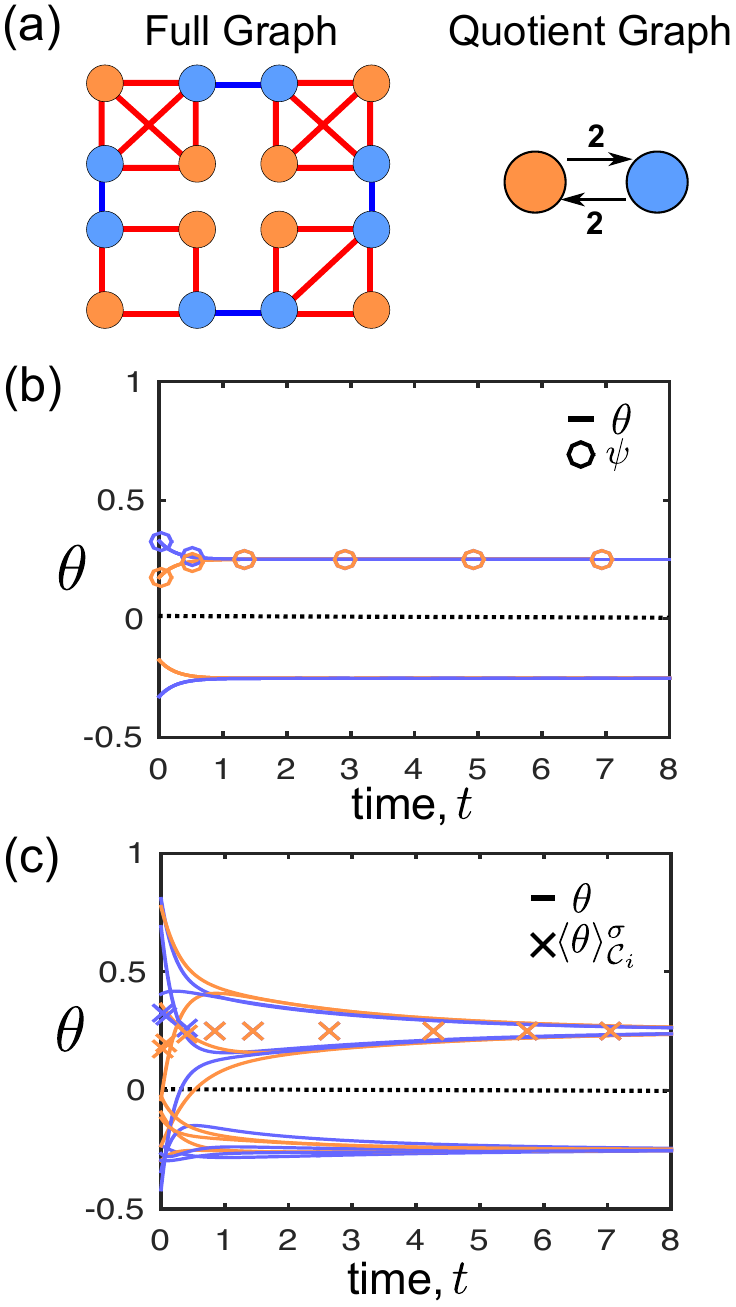}
\caption{\textbf{Bipolar cluster-synchronization on a signed graph of coupled Kuramoto oscillators with identical frequencies.} 
    \textbf{(a)} Signed graph (red links positive, blue links negative) with sEEP indicated by colors. Also shown is the associated quotient graph. 
    \textbf{(b)} For an initial condition aligned with the sEEP, the dynamics of the full system (line) is exactly equivalent to the lower dimensional Kuramoto dynamics on the quotient graph (circles) up to the sign, given by the polarization of each node. 
\textbf{(c)} 
For an initial condition ${\boldsymbol{\theta}}_0$ not too spread out on the circle, the sign-adjusted cell-averaged dynamics (crosses) governed by the quotient graph is closely aligned with the full dynamics.
The system converges to a state where nodes within a cell have phases with opposite signs.  } 
\label{fig:kuramoto_signed}
\end{figure}

While standard Kuramoto networks with positive couplings have been studied extensively~\cite{Rodrigues2016}, the literature on Kuramoto networks with both attractive (positive) and repulsive (negative) couplings is comparatively sparse, with only a handful of mean-field results~\cite{El-Ati2013}.  

Using the definition of the signed Laplacian, we write the Kuramoto model on a signed graph as:
\begin{equation}\label{eq:kuramoto_signed_large}
    \dot {\boldsymbol{\theta}} = -B_\sigma^{}\sin(B_\sigma^T{\boldsymbol{\theta}}) = -B_\sigma^{}\mathcal W(B_\sigma^T{\boldsymbol{\theta}}) B_\sigma^T  \, \boldsymbol{\theta}.
\end{equation}
Likewise, the Kuramoto dynamics on the quotient graph becomes:
\begin{flalign}
    \dot {\boldsymbol{\psi}} &= -H_\sigma^+ B \sin(B^TH_\sigma{\boldsymbol{\psi}}).
\end{flalign}
For structurally balanced signed networks, making the necessary adjustments for the 
switching equivalence $\Sigma$, 
we  then reach the same conclusions as in Section~\ref{sec:kuramoto_pos}.

In Figure~\ref{fig:kuramoto_signed}, we provide numerical examples that replicate our findings 
for signed graphs. As expected, in Figure \ref{fig:kuramoto_signed}b) we see that a bipolar cluster-synchronized solution remains invariant for all times.
Figure \ref{fig:kuramoto_signed}c) shows that for an initial condition ${\boldsymbol{\theta}}_0$ 
that is not too spread out on the unit circle, we observe numerically that the (sign adjusted) cell-averaged $\langle {\boldsymbol{\theta}} \rangle_{\mathcal{C}_i}^\sigma = H_\sigma^+ \boldsymbol{\theta} $ is well aligned with the quotient dynamics.
For non-identical intrinsic frequencies commensurate with the sEEP, the Kuramoto model converges to a final state consistent with the cells of the sEEP, yet exhibiting out-of-phase behavior within each cell due to the polarization of the nodes.

\section{Discussion}

In this paper, we have shown how to coarse-grain the dynamics of generic synchronization 
processes by using the graph-theoretical framework of external equitable partitions.
Exploiting regularities present in the underlying coupling network, 
EEPs give clusters of nodes that play an equivalent dynamical role~\cite{OClery2013}.
The resulting coarse-grained dynamics corresponds to cluster synchronization, 
in which all nodes within a cell follow the same trajectory.
Importantly, one can extend the notion of EEPs to other types of coupling schemes, as shown by our
analysis of signed networks.
In structurally balanced signed networks, we showed that each of the cells splits into two 
`out-of-phase' dynamical factions, with the same magnitude but opposite sign.

\paragraph*{\textbf{Connections with symmetry groups.}}
We have shown how our graph-theoretical approach complements
the use of symmetry groups for the analysis of synchronization dynamics over 
networks~\cite{Pecora2014,Sorrentino2016,barahona1997resonances}.
As discussed in~\citet{Pecora2014}, there exist efficient software tools to compute symmetry groups in networks~\cite{Darga2008,Group} and hence orbit partitions~\cite{Pecora2014,Sorrentino2016}. 
The more general problem of obtaining \emph{all} EEPs for a graph appears to be computationally more challenging~\cite{Kamei2013,Sorrentino2016}.
However, there exist efficient algorithms to compute EEPs centered around a node~\cite{OClery2013,Cao2013}, which can be used to characterize the dynamical influence of particular nodes on the global dynamics of the network~\cite{martini2010controllability,Egerstedt2012,OClery2013}.

\paragraph*{\textbf{Other signed coupling schemes.}}
We have chosen to consider couplings given by the signed Laplacian $L_\sigma$, as it provides a direct generalization of the standard Laplacian and has direct connections with dynamical properties.
In particular, the positive semi-definiteness of $L_\sigma$ allows us to express the Kuramoto model in terms of signed incidence matrices~\eqref{eq:kuramoto_signed_large}, thus facilitating our proof and interpretation.
However, the ideas developed here may be applied to other signed coupling schemes, provided the equivalent invariance condition to~\eqref{eq:EEP_def} can be found.
For instance, another interesting coupling is given by the
Laplacian $L_\pm=D-A$, where $A$ is a signed adjacency matrix and $D = \text{diag}(A\mathbf{1})$.
For the network in Fig.~\ref{fig:signed_EEP_consensus}a, the EEP with respect to Laplacian $L_\pm$ is almost identical to the one with respect to $L_\sigma$, except that node $8$ forms its own cell.
It is worth remarking, however, that while the algebraic characterization of such invariant partitions can still be exploited, the graph-theoretical notion of 'equitability', related to the combinatorial count of inter-cell degrees, can be lost. As many algorithms leverage such combinatorial properties to search for EEPs, this may make the presence of such invariant partitions harder to detect. Furthermore, their dynamical interpretation might be problematic in generic systems since $L_\pm$ (and other coupling matrices) will in general be indefinite, hence impacting the dynamical stability.

\paragraph*{\textbf{Relation to other synchronization notions.}} 
Because EEPs exploit the graph structure in connection with dynamics, our approach complements other methods for the analysis of synchronization. 
Here, we have concentrated on the existence and invariance of cluster-synchronized states, with some discussion of their stability in the context of the MSF. Further insights could be gained by combining our 
EEP-based analysis with other methodologies, such as the analysis of potential energy coupling landscapes or various mean-field analyses (see Refs.\cite{Doerfler2014,Arenas2008,Rodrigues2016} for overviews).
In particular, as EEPs are linked to the existence of invariant subspaces, contraction-based arguments may be fruitfully applied to global system dynamics in such networks~\cite{Moreau2004,Sepulchre2011,Russo2011,Mauroy2012}.

A particular notion of synchronization worth mentioning is that of chimera states~\cite{Panaggio2015,Abrams2004}, in which parts of the network act in unison while another parts appear unsynchronized. One may conjecture that a possible mechanism to reach such a state is to endow the underlying network with an EEP comprising one large cell and a multitude of single node cells. If, by carefully configuring the dynamics, the large cell could be made to remain stable while the single node cells follow independent trajectories, a chimera state might be obtained.

\paragraph*{\textbf{Future work.}}
Several other avenues of future work appear to be worth pursuing.
As the idea of signed networks and social balance is at the core of social network theory~\cite{Doreian1996,Cartwright1956,Heider1946}, it would be important to investigate if the bipolar cluster synchronization described here can be related to models evolving towards structural balance~\cite{Marvel2011,Traag2013}.
Following the insight by Hendrickx~\cite{Hendrickx2014} that signed opinion dynamics can be understood as a $2N$ dimensional dynamics with positive interactions, it would also be interesting to understand the symmetry requirements that an EEP implies on the lifted $2N$-dimensional graph, and whether the EEP could be used to elucidate further properties.

Another extension would be to relax the strict requirements of EEPs (e.g., by allowing minor perturbations on a graph with an EEP) in order to study how the dynamics of the system is affected. 
Generalizations of EEPs that allow different kinds of couplings (e.g., directed, time-varying~\cite{Scardovi2009, Paley2007}, delays~\cite{Papachristodoulou2010}) would also be of interest.

Finally, it is worth remarking that while we focussed here on the dynamics of 
synchronization in linear (consensus) and nonlinear processes (coupled oscillators, Kuramoto),
the concept of external equitable partitions is applicable to more general scenarios where agents interact over a graph structure.
The key ingredient is the presence of a low-dimensional invariant subspace in the coupling (spanned by the partition matrix) which can be exploited to obtain a dynamical dimensionality reduction leading to a coarse-grained system description. While EEPs have been used in consensus and control, other application areas such as ecological networks or chemical reaction networks would be worth investigating.

\section*{Acknowledgments}
We thank Karol Bacik for comments and carefully reading the manuscript.
MTS, JCD, RL acknowledge support from: FRS-FNRS; the Belgian Network DYSCO (Dynamical Systems, Control and Optimisation) funded by the Interuniversity Attraction Poles Programme initiated by the Belgian State Science Policy Office; and the ARC (Action de Recherche Concerte) on Mining and Optimization of Big Data Models funded by the Wallonia-Brussels Federation.  NO'C was funded by a Wellcome Trust Doctoral Studentship at Imperial College London during this work. YNB thanks the G. Harold \& Leila Y. Mathers Foundation. MB acknowledges support through EPSRC grants EP/I017267/1 and EP/N014529/1. 

No new data was collected in the course of this research.

\bibliography{references}
\end{document}